# The Principle of Relativity and Inertial Dragging

By

Øyvind G. Grøn

Oslo University College, Department of engineering, St.Olavs Pl. 4, 0130 Oslo, Norway
Email: Oyvind.gron@iu.hio.no

Mach's principle and the principle of relativity have been discussed by H. I. Hartman and C. Nissim-Sabat in this journal. Several phenomena were said to violate the principle of relativity as applied to rotating motion. These claims have recently been contested. However, in neither of these articles have the general relativistic phenomenon of inertial dragging been invoked, and no calculation have been offered by either side to substantiate their claims. Here I discuss the possible validity of the principle of relativity for rotating motion within the context of the general theory of relativity, and point out the significance of inertial dragging in this connection. Although my main points are of a qualitative nature, I also provide the necessary calculations to demonstrate how these points come out as consequences of the general theory of relativity.

# 1. Introduction

H. I. Hartman and C. Nissim-Sabat[1] have argued that "one cannot ascribe all pertinent observations solely to relative motion between a system and the universe". They consider an UR-scenario in which a bucket with water is at rest in a rotating universe, and a BR-scenario where the bucket rotates in a non-rotating universe and give five examples to show that these situations are not physically equivalent, i.e. that the principle of relativity is not valid for rotational motion.

When Einstein[2] presented the general theory of relativity he emphasized the importance of the general principle of relativity. In a section titled "The Need for an Extension of the Postulate of Relativity" he formulated the principle of relativity in the following way: "The laws of physics must be of such a nature that they apply to systems of reference in any kind of motion". According to the special principle of relativity an un-accelerated observer may regard himself as at rest. In the general theory an observer with any kind of motion can regard himself as at rest.

It is far from obvious that this principle is a consequence of the theory of relativity. Its possible validity depends in a decisive way upon the phenomenon of inertial dragging. C. Møller[3] has written in the following way about this: "Einstein advocated a new interpretation of the fictitious forces in accelerated systems of reference: instead of regarding them as an expression of a difference in principle between the fundamental equations in uniformly moving and accelerated systems, he considered both kinds of reference to be completely equivalent as regards the form of the fundamental equations: and the 'fictitious' forces were treated as real forces on the same footing as any other forces of nature. The reason for the occurrence in accelerated systems of reference of such peculiar forces should, according to this new idea, be sought in the circumstance that the distant masses of the fixed stars are accelerated relative to these systems of reference. The 'fictitious' forces are thus treated as a kind of gravitational force, the acceleration of the distant masses causing a 'field of gravitation' in the system of reference considered".

Ø. Grøn and E. Eriksen[4] have considered the inertial dragging effect inside a linearly accelerating spherical, massive shell and discussed the relevance of the inertial dragging effect for the possible validity of the principle of relativity for accelerated and rotating motion. Ø. Grøn and K. Vøyenli[5] have investigated the question whether the general principle of relativity is contained in the general theory of relativity, and have concluded that this is not impossible due to the inertial dragging effect, but a general proof of the validity of the general principle of relativity has not been given.

I will here discuss the relativity of rotational motion within the context of the general theory of relativity. Then the phenomenon of *inertial dragging* is essential. Its existence is a prediction of the general theory of relativity. The effect was discussed thoroughly by Lense and Thirring[6] and is therefore also called *the Lense-Thirring effect*. Its relation to Newtonian gravity is like that of magnetism in relation to the Coulomb force. Therefore it is also called *the gravitomagnetic effect*.[7]

The effect is often discussed in relation to the Kerr spacetime outside a rotating mass distribution such as the Earth. *Inertial frames*, i.e. local frames in which Newton's 1. law is valid, are dragged around by the rotation of the Earth. Due to the weakness of gravity the rotation of the inertial frames outside the Earth is extremely slow. The inertial frames use thirty million years to rotate one time around the Earth.

However the existence of this effect has recently be confirmed by measurements with Lageos II[8] and Gravity Probe B.[9]

D. R. Brill and J. M. Cohen showed that inside a rotating, massive shell the effect can be large.[10,11] They found that the inertial frames rotate with the same angular velocity as that of the shell in the limit where the Schwarzschild radius of the shell is equal to its radius. This is called *perfect dragging*. The Machian character of this result was emphasized by Brill and Cohen who wrote[10]: "A shell of matter of radius equal to its Schwarzschild radius has often been taken as an idealized cosmological model of our universe. Our result shows that in such a model there cannot be a rotation of the local inertial frame in the center relative to the large masses of the universe. In this sense our result explains why the "fixed stars" are indeed fixed in our inertial frame, and in this sense the result is consistent with Mach's principle." The phenomenon of perfect dragging has recently been demonstrated by C. Schmid in the context of expanding, flat universe models rotating slowly.[12]

A simple argument for this is the following. As noted by Brian Greene[13] (note 20, p.499): "Objects so far from us that light – or the effect of gravity – has not had sufficient time since the Big Bang to reach us, cannot influence us". The distance that light and the effect of gravity have moved since the Big Bang may be called the lookback distane, $R_0 = ct_0$ where $t_0$ is the age of the universe. WMAP-measurements[14] of the temperature variations in the cosmic microwave background radiation combined with other measurements, have shown that the age of the preferred model of our universe[15] is close to its Hubble-age, $t_H = 1/H_0$ where $H_0$ is the Hubble constant, i.e. the present value of Hubble parameter, and $t_H$ is the age the universe would have had if it had expanded with constant velocity, $t_0 = 0,996 t_H$. The WMAP-measurements also indicate that our universe is flat, i.e. that it has critical density, $\rho_{cr} = 3H_0^2/8\pi G$ where $G$ is Newton's constant of gravity. It follows that $8\pi G \rho_{cr}/3c^2 = (H_0/c)^2 \approx 1/R_0^2$. The Schwarzschild radius of the cosmic mass inside a distance $R_0$ is $R_S = 2GM/c^2 = (8\pi G \rho_{cr}/3c^2) R_0^3 \approx R_0$. Hence in our universe the Schwarzschild radius of the mass within the lookback distance is approximately equal to the lookback distance. It follows that the condition for perfect dragging is fulfilled in our universe.

In this article I shall consider the challenges presented by Hartman and Nissim-Sabat[1] to the validity of the principle of relativity of rotating motion. The main emphasis will be on how the phenomena in their challenges can be explained equally well from an UR point of view as from a BR point of view. I will provide the calculations necessary to substantiate the claim that these two points of view are physically equivalent, and hence that the principle of relativity of rotating motion is valid for the considered phenomena.

The expansion of the universe is not important for the purpose of demonstrating that there exist valid UR-explanations for phenomena originally presented from a BR-point of view. Hence the expansion of the universe will be neglected in this article. Also Brill and Cohen showed that in the first approximation spacetime is flat inside a rotating cosmic massive shell. We will therefore also neglect the curvature of spacetime on cosmic scales. These simplifications together with the argument referred to above, supporting that there is perfect dragging in our universe, will be taken advantage of by introducing a rigidly rotating frame in the flat spacetime in order to deduce the UR-explanations of the considered phenomena.

**2. Does the UR-scenario allow a non-rotating bucket?**

Hartman and Nissim-Sabat[1] note that Machian relativity requires that rotation of the universe induces rotation of a freely mounted bucket at the center of the universe.

They consider two non-rotating, coaxial buckets A and B, one of which (we do not know which) will be made to rotate. When one of the buckets (say A) has been made to rotate in a fixed universe, it will exhibit centrifugal forces, while B which remains aligned with the fixed stars, exhibit no centrifugal forces. They then write that in the UR-scenario nothing distinguishes the two buckets but a future choice. So neither of the buckets will exhibit centrifugal forces, regardless of which bucket will later be at rest in a rotating universe. Thus Machian relativity requires that all objects at the center of a rotating universe rotate, even if they are not mechanically coupled to the universe. Hence the theory of relativity for rotational motion of Mach leads to the fundamental problem that there cannot be a non-rotating bucket at the center of a rotating universe, even if the bucket and the universe are mechanically decoupled.

The solution of this problem within the context of the general theory of relativity is found by distinguishing between inertial and non-inertial reference frames and taking into account the phenomenon of perfect dragging. This phenomenon implies that *inertial frames cannot be decoupled mechanically from the universe*. Hence there cannot be an inertial, non-rotating bucket at the center of a rotating universe. But there *can* be a non-rotating freely mounted bucket at the center of the universe. This bucket in non-inertial. It has been given an angular velocity relative to the universe. However, the spin of the bucket is conserved, so it will proceed to rotate in the BR-scenario even when no force acts upon it, and in the UR-scenario it will remain at rest in the rotating universe.

**3. Centrifugal- and Coriolis acceleration as a result of inertial dragging**

We shall consider spacetime inside a cosmic shell of matter with perfect dragging of the inertial frames inside the shell. Due to the perfect dragging the cosmic mass will be non-rotating in an inertial frame, IF. Let $(T, R, \Theta, Z)$ be co-moving cylindrical coordinates in IF. In this coordinate system the line-element of the flat spacetime takes the form

$$ds^2 = -c^2 dT^2 + dR^2 + R^2 d\Theta^2 + dZ^2 . \qquad (1)$$

A non-inertial frame NIF is rotating relative to IF with an angular velocity $\omega$. An observer at rest in RF will find that the cosmic shell of mass rotates with the angular velocity $-\omega$. The coordinates $(t, r, \theta, z)$ are co-moving in NIF. The transformation between these coordinate systems is

$$t = T, \quad r = R, \quad \theta = \Theta - \omega T, \quad z = Z. \tag{2}$$

The line-element in NIF is then

$$ds^2 = -\left(1 - \frac{r^2\omega^2}{c^2}\right)c^2 dt^2 + dr^2 + r^2 d\theta^2 + dz^2 + 2r^2\omega d\theta dt. \tag{3}$$

In the IF-coordinates the non-vanishing Christoffel symbols are

$$\Gamma^R_{\Theta\Theta} = -R, \quad \Gamma^\Theta_{R\Theta} = \Gamma^\Theta_{\Theta R} = 1/R. \tag{4}$$

We now consider the motion of a particle in a plane $Z$ = constant. The acceleration of the particle in the cylindrical coordinate system is found by using the expression for the covariant derivative

$$\vec{a}_{IF} = \dot{\vec{v}} = \left(\dot{v}^i + \Gamma^i_{jk} v^j v^k\right) \vec{e}_i. \tag{5}$$

where $v^r = \dot{r}$ and $v^\theta = \dot\theta$ and the dot denotes differentiation with respect to the proper time of the particle. Inserting the Christoffel symbols (4) leads to

$$\vec{a}_{IF} = \left(\ddot{R} - R\dot\Theta^2\right)\vec{e}_R + \left(\ddot\Theta + \frac{2}{R}\dot{R}\dot\Theta\right)\vec{e}_\Theta. \tag{6}$$

In the NIF-coordinates there are additional non-vanishing Christoffel symbols,

$$\Gamma^r_{tt} = r\omega^2, \quad \Gamma^\theta_{rt} = \Gamma^\theta_{tr} = \omega/r, \quad \Gamma^\theta_{\theta t} = \Gamma^\theta_{t\theta} = r\omega, \tag{7}$$

and the covariant expression for the acceleration gives

$$\vec{a}_{RF} = \vec{a}_{IF} + \left(\Gamma^r_{\theta t}\dot\theta\dot{t} + \Gamma^r_{t\theta}\dot{t}\dot\theta\right)\vec{e}_r + \left(\Gamma^\theta_{rt}\dot{r}\dot{t} + \Gamma^\theta_{tr}\dot{t}\dot{r}\right)\vec{e}_\theta. \tag{8}$$

We now assume that the particle moves so slowly that we can put $\dot{t} = 1$. Furthermore, introducing the angular velocity vector, $\vec\omega = \omega \vec{e}_z$, an orthonormal basis with $\vec{e}_{\hat r} = \vec{e}_r$, $\vec{e}_{\hat\theta} = (1/r)\vec{e}_\theta$, a position vector $\vec{r} = r\vec{e}_{\hat r}$, a velocity vector $\vec{v} = \dot{r}\vec{e}_{\hat r} + r\dot\theta\vec{e}_{\hat\theta}$, and inserting the Christoffel symbols from eq.(7), one finds that eq.(8) takes the form

$$\vec{a}_{RF} = \vec{a}_{IF} + \vec\omega \times (\vec\omega \times \vec{r}) + 2\vec\omega \times \vec{v}. \tag{9}$$

Thus, the acceleration in NIF includes a centrifugal acceleration and a Coriolis acceleration.

These names refer to a Newtonian interpretation in which NIF is considered to be a rotating reference frame, and IF is non-rotating. Within the Newtonian theory the rotation of NIF is absolute.

However the perfect inertial dragging of the inertial frames opens up for another interpretation. An observer in NIF can regard NIF as at rest. But experiments with free particles will immediately show to him that *NIF is not an inertial reference* *frame*. The motion of the free particles does not obey Newton's 1. law. The distinction between an inertial frame and a non-inertial frame is absolute. This does not prove, however, that NIF rotates. The observer will see the heaven of stars rotating. The cosmic masses rotate in NIF. This must be taken into account when he solves Einstein's field equations to find the line element for spacetime. Using the approximation mentioned above he finds the line element (3). The extra terms in eq.(9) are due to the gravitational dragging effect. They are due to the rotation of the cosmic mass, not a rotation of NIF. According to this interpretation there is no rotation of NIF.

Newton's theory says that inertial frames are always non-rotating. It is not this way in the general theory of relativity. In the Kerr spacetime outside a rotating mass, for example, inertial frames are dragged around, and this very slow rotating motion of the inertial frames close to the Earth has now been measured with Lageos II and Gravity Probe B. Hence, *according to the general theory of relativity inertial frames may be rotating.*

**4. Rotating charged bucket**

The first challenge to the relativity of rotating motion put forward by Hartman and Nissim-Sabat[1] is the following: An electrically charged liquid in a rotating bucket produces a magnetic field that is determined by the charge density and angular velocity of the liquid. An electric field is induced in a conducting rod that is above the liquid, perpendicular to the bucket axis, and rotating with the bucket. Can one explain these magnetic and electric fields with a rotating (but presumably uncharged) universe?

This situation is similar to that of Schiff's paradox, presented in 1939[16]. We shall here consider an infinitely long, cylindrical charged shell with radius $r_1$ and surface charge density $\sigma$ rotating rigidly relative to IF with an angular velocity $\omega$. We assume that the mass per unit length of the shell is so small that the inertial dragging produced by the shell can be neglected. The magnetic field inside the cylindrical shell is

$$\hat{B} = \mu_0 \sigma \omega r_1 \quad . \tag{10}$$

In IF we have the usual expression of the electromagnetic field scalar

$$F_{\mu\nu} F^{\mu\nu} = 2\varepsilon_0 \left( c^2 \hat{B}^2 - \hat{E}^2 \right) \quad . \tag{11}$$

Since there is no electrical field we obtain (using that $\varepsilon_0 \mu_0 c^2 = 1$)

$$F_{\mu\nu}F^{\mu\nu} = 2\mu_0 \left(\sigma r_1 \omega\right)^2 \tag{12}$$

The rod rotates together with the shell. Hence in the rest frame NIF of the rod the charged shell is at rest, and it is tempting to conclude from the usual forms of Maxwell's equations that there is neither a magnetic nor an electric field inside the shell. Hence the electromagnetic fields scalar should vanish in contradiction to the prediction of an observer at rest in IF. This is essentially Schiff's paradox, and the situation was used by Feynman as an argument showing that rotational motion is absolute.[17] This was, however, presented in the Feynman Lectures of Physics in a special relativistic context, and in this context the rotational motion is indeed absolute.

In the present article the relativity of rotational motion is discussed in a general relativistic context under the assumption that we live in a universe with perfect dragging. Then Maxwell's equations in the rest frame, NIF, of the rod, do not have the same form as in an inertial frame. In NIF the Maxwell equations have the form[18]

$$\nabla \times \vec{E} + \frac{\partial \vec{B}}{\partial t} = 0 \;,\quad \nabla \cdot \vec{B} = 0$$

$$\nabla \times \left[\vec{B} - \vec{v} \times \left(\vec{E} - \vec{v} \times \vec{B}\right)\right] - \frac{\partial}{\partial t}\left(\vec{E} - \vec{v} \times \vec{B}\right) = \mu_0 \vec{j} \;,\quad \nabla \cdot \left(\vec{E} - \vec{v} \times \vec{B}\right) = \frac{\rho}{\varepsilon_0} \tag{13}$$

where $\vec{v} = \vec{\omega} \times \vec{r}$ and $\vec{j}$ is the current density. Schiff notes that the extra terms in eq.(13) are due to the gravitational action of the rotating distant masses in NIF.

In the present case the fields are stationary and there are no currents in NIF. Hence, the equations reduce to

$$\nabla \times \vec{E} = 0 \;,\quad \nabla \cdot \vec{B} = 0 \;,\quad \nabla \times \left[\vec{B} - \frac{\vec{v}}{c^2} \times \left(\vec{E} - \vec{v} \times \vec{B}\right)\right] = 0 \;,\quad \nabla \cdot \left(\vec{E} - \vec{v} \times \vec{B}\right) = \frac{\rho}{\varepsilon_0} \;. \tag{14}$$

From the cylindrical symmetry of the problem it follows that the only non-vanishing components of $\vec{E}$ and $\vec{B}$ are $E_r(r)$ and $B_z(r)$. The two first equations in (14) are satisfied by all fields of this form. Since $\vec{v} \cdot \vec{B} = 0$, the vector identity

$$\vec{a} \times \left(\vec{b} \times \vec{c}\right) = \left(\vec{a} \cdot c\right)\vec{b} - \left(\vec{a} \cdot \vec{b}\right)\vec{c} \tag{15}$$

gives

$$\vec{v} \times \left(\vec{v} \times \vec{B}\right) = -v^2 \vec{B} \;. \tag{16}$$

The third of eqs.(14) then reduces to

$$\nabla \times \left[\left(1-\frac{v^2}{c^2}\right)\vec{B} - \frac{\vec{v}}{c^2} \times \vec{E}\right] = 0 , \qquad (17)$$

which in the present case leads to

$$\frac{d}{dr}\left[\left(1-\frac{r^2\omega^2}{c^2}\right)B + \frac{r\omega}{c^2}E\right] = 0 . \qquad (18)$$

Integration gives

$$\left(1-\frac{r^2\omega^2}{c^2}\right)B + \frac{r\omega}{c^2}E = B(0) . \qquad (19)$$

Inside the shell the last of eqs.(14) reduce to

$$\nabla \cdot (\vec{E} - \vec{v} \times \vec{B}) = 0 , \qquad (20)$$

or

$$\frac{d}{dr}(E - r\omega B) = 0 . \qquad (21)$$

Integration gives

$$E - r\omega B = E(0) . \qquad (22)$$

Due to the cylindrical symmetry the electrical field vanishes at the axis. Hence $E(0) = 0$ and

$$E = r\omega B . \qquad (23)$$

Inserting this into eq.(19) gives $B = B(0)$, i.e. the magnetic field is homogeneous inside the cylinder and equal to that in IF,

$$B = \mu_0 \sigma \omega r_1 . \qquad (24)$$

Using eqs.(6) and (7) of ref.12 we find that in NIF the expression (11) is replaced by

$$F_{\mu\nu}F^{\mu\nu} = 2\varepsilon_0\left[(c^2 - r^2\omega^2)B^2 + E^2\right] . \qquad (25)$$

Inserting eq.(23) we see that the two last terms inside the bracket cancel, and using eq.(24) we finally arrive at

$$F_{\mu\nu}F^{\mu\nu} = 2\mu_0(\sigma r_1 \omega)^2 \qquad (26)$$

in accordance with eq.(12).

A. Bhadra and S. C. Das[19] have written some comments to the challenges of Hartman and Nissim-Sabat where they defend the validity of the principle of relativity for rotational motion. However they give no calculations, and concerning the present challenge they only write: "In both the UR and BR cases, the liquid is rotating with respect to the observer tied to the universe. Hence the same derivation is applicable for the UR and BR cases."

The important point as to the validity of the principle of relativity for rotating motion is however, whether the physical situation has a valid explanation from the UR point of view. Can one explain the potential difference along the rod assuming that the rod is at rest and the universe rotates around it? And if one can, what is the explanation? In IF the potential difference is due to two facts: firstly that the rotating charge produces a magnetic field, and secondly that induction produces an electrical field and hence a potential difference along the rotating rod.

Neither of these phenomena exists in the common rest frame of the fluid and the rod. In this frame the magnetic field along the axis of the bucket and the radial electrical field are due to modifications of Maxwell's equations that come from the inertial dragging due to the rotating cosmic masses. Newton's 1. law is not obeyed by a free particle in NIF. Hence although the observers in NIF can consider this frame as at rest, they must agree that it is not an inertial frame. Therefore additional terms appear in Maxwell's equations. The magnetic and electrical fields inside the charged shell, which is at rest in NIF, are consequences of these terms.

**5. Radiation from a rotating charged bucket**

The second challenge to the principle of relativity for rotating motion given by Hartman and Nissim-Sabat[1] is: "A rotating charged liquid radiates electromagnetic radiation that carries energy away. So one must do work to keep the bucket rotating at a constant speed. Can one explain the radiation flux and the work done on the bucket in the UR-scenario?"

In order to simplify the analysis we shall consider one point charge in the liquid, neglecting the others. The charge moves along a circular path in the inertial frame IF and is permanently at rest in NIF.

It is important to note that even if the existence of electromagnetic radiation from a charge is invariant against a Lorentz transformation, it is not invariant against a transformation involving accelerated and rotating reference frames. This was first shown by M. Kretzschmar and W. Fugmann[20,21], and later re-derived in connection with electromagnetism in uniformly accelerated reference frames by T. Hirayama[22] and in a different way by E. Eriksen and Ø. Grøn[23]. It was shown that Larmor's formula for the power radiated by a point charge must be generalized in order to be applicable to non-inertial reference frames. The power radiated by a charge depends upon the square of the charge's acceleration *relative to the reference frame*. Hence, an observer at rest in NIF will not detect any radiation from the charge which is at rest in NIF although an observer in IF detects radiation from it.

However the tangential force that must be used in IF in order that the charge shall move along a circular path with constant velocity[24] is present in NIF as well. The force acts in the direction of motion of the charge in IF. In NIF the cosmic masses are observed to move in the opposite direction, and the inertial frames, i.e. IF, moves in this direction. The tangential force is then needed in order to keep the charge at rest. But this force does no work in NIF.

The answer to Hartman and Nissim-Sabat's second challenge is therefore: There is no radiation flux in the UR-scenario and no work is performed on the bucket.

In the third challenge they write: "If work is not provided, the bucket spin slows down and the kinetic energy lost equals the energy radiated. With UR, the angular velocity of the universe decreases at the same rate as the angular velocity of the bucket with BR, but the kinetic energy lost is much larger."

The energy budget in the rest frame of the liquid is different from that in IF, since there is no radiation in NIF. Also, from the BR-point of view one would say that the rest frame of the liquid, NIF, is not only rotating, but it has an angular acceleration as well.

In order to defend the relativity of rotating motion for this case one has to solve Einstein's field equations inside a massive, cosmic shell which rotates with a decreasing angular velocity. As far as I know this has not been done, and I will here only conjecture that there will be a dragging field adapted to the decreasing angular velocity of the cosmic masses as observed in NIF. Hence in NIF there will be a decreasing centrifugal field. This means that according to the UR-point of view the (negative) potential energy (with zero at the axis) of the cosmic masses increases. And since no work is performed the increase of potential energy will equal the decrease of kinetic energy.

**6. The Sagnac experiment**

The fourth challenge of Hartman and Nissim-Sabat[1] concerns the Sagnac experiment[25]. The experiment showed that there is a fringe shift between co-rotating and counter-rotating light beams travelling in a rotating polygon. All special relativistic effects are of second order in $v/c$. But the fringe shift is of first order in $r\omega/c$. Therefore the Sagnac experiment was interpreted within the context of the special theory of relativity to show that rotational motion is absolute. According to the BR-point of view the fringe shift is due to the motion of the detector along a circular path during the time the light travels from the emitter to the detector.

Again, the situation is interpreted in a different way in the general theory of relativity. The light follows a null-geodesic circular curve with constant *r* and *z*. From eq.(3) we then have

$$c_N^2 + 2r\omega c_N - (c^2 - r^2\omega^2) = 0 \qquad (27)$$

where $c_N = r\, d\theta/dt$ is the coordinate velocity of the light in NIF. The solutions of this equation are

$$c_{N\pm} = -r\omega \pm c \qquad (28)$$

This shows that the coordinate velocity of light is anisotropic in NIF. The difference of the travelling time for light travelling around a circle in opposite directions is

$$\Delta t = \frac{2\pi r}{c - r\omega} - \frac{2\pi r}{c + r\omega} = \gamma^2 \frac{4\pi r^2 \omega}{c^2} \qquad (29)$$

This accounts for the fringe shift as referred to NIF and gives the same, invariant result as that in IF.

However, the explanation of the fringe shift is different from that in IF. It is due to the anisotropy of the velocity of light, which is a signature showing that NIF is not an inertial frame. Nevertheless the observers in NIF may perfectly well explain this experiment from the UR-point of view. Then the inertial frames are considered to rotate due to the dragging effect of the rotating cosmic masses.

## 7. Astronomical observations

The fifth challenge to the relativity of rotational motion presented by Hartman and Nissim-Sabat[1] is concerned with stellar aberration. They claim that astronomical observations contradict Machian relativity. In the UR-scenario a star sufficiently far away from the observer is said to move with a superluminal velocity due to the rotating motion of the universe. Hence, light from the star generates a shock wave producing a Cherenkov effect. No such phenomenon results in the BR-scenario. Their conclusion is that astronomical observations show that the BR-scenario and the UR-scenario are physically different, and hence the relativity of rotating motion is disproved.

The general theory of relativity provides an interesting solution to the challenge. The essential point is that the concept of space is different in this theory than in Newtonian physics or the special theory of relativity. This has recently been discussed in connection with the expansion of the universe[26]. In cosmology space is defined by a field of free particles, i.e. of local inertial frames. The theory allows superluminal expansion velocity, and this velocity does not produce any sort of shock wave.

It is peculiar velocities, i.e. motion *through* space that is restricted to be less than the velocity of light, in the sense that material particles have world lines inside the future light cone of the emitter event.

Considering the motion of the stars in the rotating universe their peculiar velocities have been assumed to vanish in the discussion of Hartman and Nissim-Sabat[1]. Hence, according to the general theory of relativity, their superluminal velocity due to the rotation of the universe, generate no shock wave. Under the assumption of perfect dragging no observable astrophysical phenomena will appear in the UR-scenario other than those that are present in the BR-scenario.

## 8. Conclusion

In the present article we have considered several challenges to the validity of the principle of relativity for rotating motion recently raised by Hartman and Nissim-Sabat[1]. Assuming perfect dragging in our universe it has been shown how the mentioned observations may be explained both in the BR-scenario where the bucket rotates in a non-rotating universe and in the UR-scenario in which a bucket with water is at rest in a rotating universe.

The observed phenomena have different explanations in the BR- and UR-scenario. Centrifugal- and Coriolis forces result from the tendency of free particles to move along straight paths in inertial frames. In the BR-scenario the non-inertial reference particles of a frame rotating counter clockwise, turn to the left. In relation to these reference particles, free particles turn to the right, which is why free particles instantaneously at rest in NIF accelerate outwards (centrifugal acceleration) and moving particles accelerate to the right (Coriolis acceleration). From the UR-point of view NIF is at rest in a rotating universe, and there is perfect dragging of free particles, which is a gravitational effect of the rotating cosmic masses. This causes free particles at rest to accelerate outwards and moving particles to accelerate to the right.

In the BR-scenario a rotating charged fluid produces a magnetic field, and induction then produces a radial electric field causing a voltage over a rod co-moving with the fluid. In the UR-scenario the fluid and the rod are at rest in a rotating universe. The rotating cosmic masses causes additional terms in Maxwell's equations, not present in inertial frames. The solution of these equations inside a charged shell at rest shows that there exist a magnetic field along the symmetry axis of the cylindrical shell and a radial electric field causing the same voltage in NIF that is measured in IF.

A charge moving circularly in flat spacetime emits electromagnetic radiation. Due to the radiation reaction force the charge must be acted upon by a tangential force in order to move with constant velocity. The work performed by this force accounts for the emitted radiation. In NIF the charge is permanently at rest and does not emit radiation. Nevertheless the same force acts. It is needed to keep the charge at rest. Without this force the charge would start moving due to the inertial dragging caused by the rotating cosmic masses.

If the external force does not act, the tangential velocity of the charge would decrease. In this case the rotational velocity of the cosmic masses would decrease in NIF, and there would be a huge decrease of kinetic energy of the cosmic masses although no external force acts upon it. This seeming energy paradox is solved by the centrifugal field in NIF. This gets weaker since the rotational velocity of the comic masses decreases. Hence, the potential energy of the cosmic mass gets an increase equal to its loss in kinetic energy during the motion.

In the usual BR-scenario the fringe shift in the Sagnac experiment is due to the motion of the receiver during the time the light moves from the emitter to the receiver, because of the rotation of the apparatus. In the UR-scenario there is no rotation of the apparatus. The velocity of light is isotropic in the field of local inertial frames. And they are dragged by the rotating cosmic mass. Hence the velocity of light is different in the direction of motion of the inertial frames and in the opposite direction. This produces the observed fringe shift in the Sagnac experiment in the UR-scenario.

Far away stars move with superluminal velocity in the UR-scenario. One might think that light emitted from these stars produces a bow shock giving rise to a Cherenkov effect, making the UR-scenario different from the BR-scenario. However this is not the case. The reason is that the field of inertial frames define the cosmic space. The rotating motion of the inertial frames is like the expansion velocity of the Hubble flow in expanding universe models. Superluminal velocity of the local inertial frames is allowed, and light emitted from these frames produces no Cherenkov effect. I conjecture that there are no observable astrophysical effects in the UR-scenario that is not also present in the BR-scenario.

The result of the present investigation is that all the challenges to the validity of the principle of relativity for rotating motion can be dealt with within the context of the general theory of relativity. Such challenges are interesting because *the explanation of physical phenomena depend upon the frame of reference that is used*. Many phenomena have explanations that we are used to only in the BR-scenario, i.e. from the point of view that a system rotates in a non-rotating universe. Much can be learned about the physical contents of the general theory of relativity by trying to explain the same phenomena in the UR-scenario, i.e. from the point of view of a non-rotating system in a rotating universe.

Finally one may note that the phenomenon of perfect dragging is essential for the possibility of explaining the phenomena considered by Hartman and Nissim-Sabat[1] from the UR-point of view. As far as this phenomenon can be proved, it seems that one has made an important step towards demonstrating the validity of the general principle of relativity within the general theory of relativity.